\documentclass[letterpaper,aps,prl,twocolumn,floatfix,amsfonts,amssymb]{revtex4-1}
\usepackage{graphicx}
\usepackage{amsmath}
\usepackage{hyperref}
\usepackage[dvipsnames]{xcolor}

\hypersetup{
  colorlinks   = true, 
  urlcolor     = Blue, 
  linkcolor    = Blue, 
  citecolor   = Blue 
}

\usepackage{color}

\begin{document}
\title{Exact nonequilibrium transport in the topological Kondo effect}
\author{B. B\'eri}
\affiliation{School of Physics \& Astronomy, University of Birmingham, Edgbaston, Birmingham, B15 2TT, United Kingdom}
\date{October 2016}
\begin{abstract}
A leading candidate for experimental confirmation of the non-local quantum dynamics of Majorana fermions is the topological Kondo effect, predicted for  mesoscopic superconducting islands connected to metallic leads. We identify an anisotropic, Toulouse-like, limit of the topological Kondo problem where the full nonequilibrium conductance and shot noise can be calculated exactly. Near the Kondo fixed point, we find novel asymptotic features including a universal conductance scaling function, and fractional charge quantisation observable via the Fano factor. In the universal regime, our results apply for generic anisotropy and even away from the Kondo limit as long as the system supports an emergent topological Kondo fixed point. Our approach thus provides key new qualitative insights and exact expressions for quantitative comparisons to future experimental data.
\end{abstract}
\maketitle

Majorana fermions are exotic quasiparticles arising in topological superconductor structures \cite{KitaevPU2001,*FuKane08,*LutchynPRL2010,*oreg2010helical,*AliMajrev,*BeeMajRev}. In their most often studied form they are spatially localised modes which, when far apart, have zero energy and encode ordinary fermions in a nonlocal manner. This gives rise to a topologically degenerate ground state subspace, in which the nonlocal fermions are proposed as topological qubits for fault tolerant quantum computation \cite{kitaev2003fault,HasslerNJoP2010,*HeckNJP2012,*HyartPRB2013}. 

Of significant current interest, both due to the proposed Majorana signatures they support \cite{Futelep,HeckPRB2011,*HutzenPLR2012,*HeckPRB2016,*VuikNJP2016,BeriTK,MulticoulMaj,*MajKlein,ABET1,zazunov2014transport,GalpinPRB14,MichaeliarXiv2016,*HerviouarXiv2016} and a number of specific computational schemes they are expected to enable \cite{HasslerNJoP2010,*HeckNJP2012,*HyartPRB2013}, are Majorana devices based on mesoscopic superconductor islands where charging effects are significant. After finding experimental signatures consistent with the zero energy nature of Majorana fermions \cite{Mourik25052012,*NadjPerge2014,*das2012zero}, turning to such mesoscopic devices led to the first results \cite{AlbrechtNat2016} suggestive of the nonlocality of the Majorana based fermions in the form of electron teleportation \cite{Futelep}, though possible non-Majorana based explanations for the observations were noted to exist \cite{SauPRBnonlocality}. 

\begin{figure}
\includegraphics[trim=0 2 0 0,clip,width=0.85\columnwidth]{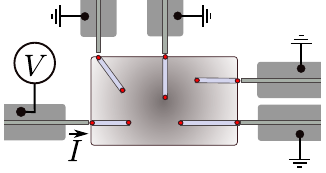}
\caption{Sketch of an $M=5$ topological Kondo setup: a mesoscopic  superconducting island hosting Majorana fermions (red dots), coupled to $M$ leads of conduction electrons. We focus on the conductance $G=\frac{\partial I}{\partial V}$ and zero-frequency shot noise $P$ associated with the current $I$ in one of the leads when it is biased with voltage $V$ with respect to the rest.}
\label{fig:kondosetup}
\end{figure}

A compelling signature of the Majorana nonlocality and of topological qubits would be the observation of the so-called topological Kondo effect \cite{BeriTK,MulticoulMaj,*MajKlein}, predicted to arise in mesoscopic charging dominated devices with $M\geq 3$ leads connected to $M$ Majorana fermions (an example with $M=5$ is shown in Fig.~\ref{fig:kondosetup}). In this effect, topological qubits play the role of a nonlocal SO($M$) ``impurity spin" for the Kondo effect, and lead to signatures that include a conductance enhancement with non-Fermi liquid low energy features (e.g., fractionally quantised power laws and zero energy conductance). In a minimal, $M=3$ lead device these features can be turned off by decoupling any one of the leads, providing an additional, highly qualitative handle on the effect.

Here we describe an exact approach for calculating the nonequilibrium conductance and shot noise in topological Kondo systems, focusing on the universal regime below the Kondo temperature $T_K$, the sole energy scale characterising the low energy physics. For the conductance, we provide the combined temperature $T$ and voltage $V$ dependence, which, even in terms of low energy asymptotes,  was unavailable so far. In fact, beyond asymptotes, the only conductance study was the numerical simulation of the $T$ dependent linear AC regime in Ref.~\onlinecite{GalpinPRB14}. Our exact results give access to the complementary nonlinear DC behaviour.
For the shot noise we focus on the zero temperature DC regime. In addition to the exact results, our approach will be shown to provide new physical insights, 
uncovering the emergence of a quantised fractional charge $e^*=\frac{2(M-1)}{M}e$ observable in the Fano factor.

The universality of the regime below $T_K$ is meant
in the sense numerically \cite{Krishna-murthyPRB1980,*CostiJPC1994} and experimentally \cite{Goldhaber-GordonPRL1998} demonstrated for conventional Kondo devices: it is expected to describe the low energy physics in a broad range of settings including, beyond the idealised Kondo case $\Gamma_t\ll \Delta E$ and $T_{K}\sim\Delta Ee^{-\Delta E/\Gamma_{t}}$ (with $\Gamma_t$ the typical Majorana level broadening and $\Delta E$ the minimum energy for changing the island charge by $\pm e$), also high $T_K$ devices with $\Gamma_t\gtrsim \Delta E$ and,
 at least asymptotically, even $\Gamma_t\gg \Delta E$ systems near charge degeneracy as suggested by their topological Kondo low energy physics \cite{MichaeliarXiv2016}.

The key technical innovation behind our results is a mapping between the topological Kondo problem in a suitably chosen anisotropic limit, and  the (massless) boundary sine Gordon (BSG) model in its form as impurity backscattering in a repulsive Luttinger liquid. Motivated by the existence of analogous mappings between BSG(-equivalent) and more conventional anisotropic Kondo models \cite{ToulousePRB1970,*EK, *FabGog,YiKane,*YiQBM}, we call this anisotropic limit ``Toulouse limit". The feature that makes our Toulouse limit well suited for transport calculations is that 
it maintains a clear link to the particle number
of an (arbitrarily chosen) lead $M$ and 
the overall particle number
of the rest of the leads, thus 
allowing us to exploit exact BSG results pioneered by Fendley, Ludwig and Saleur \cite{FendleyPRL95,*FendleyPRB95,FendleyNoisePRL95}. 
As in more conventional cases, universality implies that the specific anisotropy involved in our Toulouse limit is not a limitation  as long as energies sufficiently below $T_K$ are probed. In fact, universality and the emergent isotropy of the Kondo problem at low energies means that the observables  calculable from the Toulouse limit (see Fig.~\ref{fig:kondosetup}) have an extended scope, e.g., the conductance, upon rescaling, is expected to inform on the whole low energy conductance tensor.

We now turn to describing the Toulouse limit and its transport applications. Our starting point is the topological Kondo Hamiltonian \cite{BeriTK}, describing
Majorana assisted hopping between leads of conduction electrons, 
\begin{equation}
H=\sum_{j}H_{0j}+\sum_{j\neq k=1}^{M}\lambda_{jk}\gamma_{j}\gamma_{k}\psi_{k}^{\dagger}(0)\psi_{j}(0).
\label{eq:TKfermionform}
\end{equation}
Here $\lambda_{jk}=\lambda_{kj}>0$ and compared to Ref.~\onlinecite{BeriTK} we omitted a term inoperative in the universal regime. In Eq.~\eqref{eq:TKfermionform},  $H_{0j}$ is the Hamiltonian of conduction elecrons $\psi_j$ in half-infinite lead $j$ defined for $x<0$, with $x=0$ being the tunneling point to Majorana fermion $\gamma_j$. 
The reason Eq.~\eqref{eq:TKfermionform} is a Kondo problem is because  $\gamma_j\gamma_k$ realise topological qubit SO($M$) ``impurity spin" operators, whilst $\psi_{k}^{\dagger}\psi_{j}$ can be arranged to form conduction electron SO($M$) ``spin" densities \cite{BeriTK}. We emphasise that, due to the nonlocality of Majorana topological qubits, these ``spin" objects are highly nonlocal. The resulting nonlocal transport phenomena [e.g., the SO($M$) to SO($M-1$) switch in  Kondo features upon decoupling any one of the leads] are direct signatures of the Majorana nonlocality. 

In what follows, we take $H_{0j}$ to be noninteracting, describing, e.g., 
Fermi liquid electrodes coupled to Majorana fermions via short wire segments (quantum point contacts) \cite{ChamonFradkin};
however, longer, interacting wires can also be straightforwardly described \cite{MulticoulMaj,*MajKlein}.
For devices satisfying the Kondo model condition $\Gamma_t\ll \Delta E$,  $\lambda_{jk}\sim t_jt_k/ \Delta E$ where $t_j$ is the lead-Majorana tunnel amplitude (chosen to be positive without loss of generality), in terms of which $\Gamma_{t}\sim\nu\sum_{j}t_{j}^{2}$ with $\nu$ the lead density of states. For devices with $\Gamma_t\gtrsim \Delta E$, if they display topological Kondo physics at low energies, Eq.~\eqref{eq:TKfermionform} may be used to capture the universal regime  so long as $T_K$ itself is used as the reference energy scale (though here the relation of $T_K$ to microscopics is different). 

To formulate the Toulouse limit, 
it is profitable to note that Majorana assisted hopping problems can be effectively cast into bosonic tunneling \cite{MulticoulMaj,*MajKlein,TsvelikPRL2013,*crampe2013quantum,MichaeliarXiv2016} via bosonisation \cite{Giabook}. In terms of this, the leads have free boson Hamiltonian  $H_{0j}=\frac{\hbar v_F}{8\pi}\int dx(\partial_{x}\theta_{j})^{2}+(\partial_{x}\varphi_{j})^{2}$ with $v_F$ the Fermi velocity, $\varphi_j$ encoding the 
density $\rho_j =\frac{\partial_x\varphi_j}{2\pi}$ and $\theta_j$ being its canonical conjugate, $[\partial_{x}\varphi_{j}(x),\theta_{j^{\prime}}(x^{\prime})]=4\pi i\delta(x-x^{\prime})\delta_{jj^{\prime}}$.  
At weak coupling ($\nu\lambda_{jk}\ll 1$),
the fields obey boundary conditions $\varphi_j(0)=(\partial_x \theta_j)(0)=0$ and the electron operator, at the tunneling point is $\psi_j(0)=\frac{i}{\sqrt{a}}\Gamma_je^{i\theta_j/2}$ ($a$ is the short distance cutoff). Here $\Gamma_j$ 
is a Klein factor, also a Majorana fermion. A key feature of Majorana fermions 
is that for different $j$ the products 
$i\gamma_j\Gamma_j=\pm 1$ mutually commute and thus effectively cancel from the problem, leaving behind the fully bosonic Hamiltonian \cite{MajKlein}
\begin{equation}
H=\sum_{j}H_{0j}-2\sum_{j<k}\lambda_{jk}\cos\frac{\theta_{k}(0)-\theta_{j}(0)}{2},
\label{eq:TKbosonform}
\end{equation}
where $a$ has been absorbed into $\lambda_{jk}$.

The Toulouse limit consists of sending $\lambda_{j,k\neq M}\rightarrow\infty$ in Eq.~\eqref{eq:TKbosonform}. [Note that this is not equivalent to $\lambda_{j,k\neq M}\rightarrow\infty$ in Eq.~\eqref{eq:TKfermionform}; indeed the Kondo fixed point is $\lambda_{jk}\rightarrow \infty$ for  Eq.~\eqref{eq:TKbosonform} while it is at intermediate coupling in terms of Eq.~\eqref{eq:TKfermionform}.] 
We will now show that in this limit Eq.~\eqref{eq:TKbosonform} can be transformed into backscattering in a repulsive Luttinger liquid, Eq.~\eqref{eq:Luttwire} below.
We start with an orthogonal rotation  inspired by works  on quantum Brownian motion \cite{YiKane,*YiQBM}:  
we decompose $(\varphi_1,\ldots, \varphi_{M-1})$ and $(\theta_1,\ldots,\theta_{M-1})$ into $\theta_c=\frac{\sum_{j=1}^{M-1}\theta_j}{\sqrt{M-1}}$,  $\varphi_c=\frac{\sum_{j=1}^{M-1}\varphi_j}{\sqrt{M-1}}$ associated with the overall charge in leads $1,\ldots,M-1$; and ``spin" fields $\tilde{\theta}_j$, $\tilde{\varphi}_j$, $j=1\ldots M-2$ which are the components in directions orthogonal to $(1,\ldots,1)$. This is useful because the $\lambda_{j,k\neq M}$ terms involve only $\tilde{\theta}_j$. 
The Toulouse limit amounts to new spin field boundary conditions, pinning $\tilde{\theta}_j$ so that the $\lambda_{j,k\neq M}$ terms are minimised. 
With these terms effectively rendered constant, one gets
\begin{equation}
H=H_{0M}+H_{c}+\lambda\cos\frac{\theta_{M}-\theta_{c}/\sqrt{g_{t}}}{2}+H_{s}\label{eq:HuneqalLutt}.\end{equation}
Here $H_{c,s}$ are free boson Hamiltonians for the charge and spin fields, $g_t\!=\!M\!-\!1$, $\lambda\!=\!\sum_{j=1}^{M-1}\!2\lambda_{Mj}$, and we absorbed a phase in $\theta_c$. 
Note that Eq.~\eqref{eq:HuneqalLutt} does not contain perturbations tunneling $\tilde{\theta}_j$ between pinning positions, even though the new boundary conditions are compatible with these (see e.g., Ref.~\onlinecite{MajKlein}). The formal $\lambda_{j,k\neq M}\rightarrow\infty$  limit expresses that these perturbations are set to zero: their absence, and thus the decoupling of the $M-1$ lead spin sector, is part of the definition of the Toulouse limit.

Apart from $H_s$, Eq.~\eqref{eq:HuneqalLutt} is equivalent to tunneling between two half-infinite wires: the physical $M$-th lead and a fictitious lead accounting for the overall density $\rho_c\!=\!\frac{\sqrt{g_t}}{2\pi}\partial_x \phi_c$ of the other  
leads. This fictitious lead, as indicated by the Luttinger parameter $g_t\!>\!1$,  has attractive interactions. A similar correspondence was alluded to by Nayak \textit{et al.} in a seminal work on multilead tunneling \cite{Nayak99}. Here we establish the $\lambda_{j,k\neq M}\!\rightarrow\!\infty$ Toulouse limit as the concrete framework where this correspondence arises, which is our first key result.

To make direct contact with the backscattering model, 
we make a final sequence of transformations \cite{ChamonFradkin}: 
we first join up the left and right moving modes of the half-infinite wires to ``unfold" \cite{Nayak99} them into infinite wires supporting right moving modes $\chi_M$, $\chi_c$ and $\tilde{\chi}_j$ with $[\chi_{j}(x),\chi_{j^{\prime}}(x^{\prime})]=i\pi\text{sgn}(x-x^{\prime})\delta_{jj^{\prime}}$. We then perform an orthogonal rotation \cite{ChamonFradkin}
\begin{equation}
\left(\begin{array}{c}
\phi_{R1}\\
\phi_{R2}
\end{array}\right)=\left(\begin{array}{cc}
\frac{\sqrt{M-1}+1}{\sqrt{2M}} & \frac{\sqrt{M-1}-1}{\sqrt{2M}}\\
-\frac{\sqrt{M-1}-1}{\sqrt{2M}} & \frac{\sqrt{M-1}+1}{\sqrt{2M}}
\end{array}\right)\left(\begin{array}{c}
\chi_{M}\\
\chi_{c}
\end{array}\right)
\label{eq:CMrot}\end{equation}
and finally introduce the right and left movers, $\phi_R(x)=\phi_{R1}(x)$, $\phi_L(x)=-\phi_{R2}(-x)$, respectively. 
After these steps, Eq.~\eqref{eq:HuneqalLutt} becomes $H=H_{\text{Lutt}}+H_s$ where 
\begin{equation}
H_{\text{Lutt}}=H_{L}+H_{R}+\lambda\cos\left(\sqrt{g_{b}}[\phi_{R}(0)+\phi_{L}(0)]\right),
\label{eq:Luttwire}
\end{equation}
with $H_{\alpha}=\frac{\hbar v_{F}}{4\pi}\int dx(\partial_{x}\phi_{\alpha})^{2}$, $g_b=\frac{M}{2(M-1)}$, and $\alpha=R,L$ mover 
densities $\rho_{\alpha}\!=\!\frac{\sqrt{g_{b}}}{2\pi}\partial_{x}\phi_{\alpha}$. 
The particle numbers $Q_\beta=\int \rho_\beta(x)dx$ ($\beta=R,L,M,c$) are directly linked via Eq.~\eqref{eq:CMrot}, as will be utilised below.
$H_\text{Lutt}$ is the problem of backscattering in a Luttinger liquid \cite{KFluttlett,*KFluttPRB,FendleyPRL95}. The Luttinger parameter in this picture is $g_b<1$, corresponding to repulsive interactions. 
Given the equivalence of the backscattering and BSG models, Eq.~\eqref{eq:Luttwire} completes the transformation. 

The above mapping between topological Kondo and BSG models already suggests that 
the latter may capture the universal regime of topological Kondo systems. 
There is however further consideration needed to support this expectation: so far what we have is a correspondence for weak $\lambda$. This regime  describes only the high energy features of Eq.~\eqref{eq:Luttwire};  in terms of the standard renormalisation group (RG) argument \cite{KFluttlett}, this is because the backscattering term has scaling dimension $g_b<1$ and is thus a relevant perturbation. If Eq.~\eqref{eq:Luttwire} is to capture the low energy Kondo physics, this must be via the RG flow towards strong coupling. We now make it plausible that this indeed happens, offering several pieces of evidence. 

The first 
is the boundary entropy $S_b$, which in our context is the difference between the ground state entropy of $\lambda_{jM}\neq 0$ and $\lambda_{jM}=0$ systems. For BSG models, this is a known function of $g_b$ \cite{FSW1994}. It is given by $S_{b}\!=\!\ln\!\sqrt{g_{b}}\!=\!\ln\!\sqrt{\frac{M}{2(M-1)}}$, which, remarkably, is precisely the difference between boundary entropies of $M$ and $M\!-\!1$ lead  topological Kondo fixed points \cite{ABET2}. This indicates that the 
RG flow endpoints of the BSG model
indeed have the intended topological Kondo interpretation.

Further evidence is given by the conductance near the strong coupling fixed point, most transparently analysed in terms of Eq.~\eqref{eq:HuneqalLutt}. 
At the fixed point, the conductance between the half infinite wires of Eq.~\eqref{eq:HuneqalLutt} is $G_0=\frac{2e^2}{h}\frac{M-1}{M}$ \cite{ChamonFradkin,ChangYu12}. The same value holds for the topological Kondo fixed point. Furthermore, for the low energy correction $\delta G=G_0-G$, the leading power law $\delta G\sim E^{2/g_b-2}$ arising when there is only one infrared energy scale $E$ of interest (e.g., $E=eV\gg T$ or $E=T\gg eV$) also matches because the leading strong coupling scaling dimension $\frac{1}{g_b}=\frac{2(M-1)}{M}$ of the junction \cite{ChangYu12} is the same as for the topological Kondo fixed point \cite{BeriTK, MajKlein, Toulpertfn}.

With this preparation, we can now turn to describing how the exact results for the BSG model can be exploited to calculate topological Kondo transport properties. 
This will provide a framework analogous to that of Schiller and Hershfield \cite{SchillerHershfield1,*SchillerHershfield2} who leveraged the Toulouse limit for transport in more conventional Kondo systems.
The BSG model is well known to have a crossover energy scale $T_K\sim\lambda^{1/(1-g_b)}$ separating high and low energies \cite{FendleyPRL95}; we identify this with the Kondo temperature. 
(Note that the value of $T_K$ in the Toulouse limit depends on which lead is chosen as the $M$-th; this, however, does not affect the universal features that emerge upon using $T_K$ as the reference scale.)
The tunneling from or to lead $M$ corresponds to backscattering in Eq.~\eqref{eq:Luttwire} therefore the current $I$ in this lead will correspond to the backscattering current $I_b$ in terms of Eq.~\eqref{eq:Luttwire}.  
For the precise correspondence, one has to relate $\Delta Q = Q_M-Q_c$  and  $\Delta Q_{RL} = Q_R-Q_L$.
These quantities set both the currents 
$I=\frac{e}{2}\partial_t \Delta Q$ and $I_b=\frac{e}{2}\partial_t \Delta Q_{RL}$ and how voltage bias enters. Using Eq.~\eqref{eq:CMrot}, we find $\Delta Q=\frac{\Delta Q_{RL}}{g_{b}}+\frac{2-M}{\sqrt{2Mg_{b}}}Q^{\text{tot}}$, where $Q^{\text{tot}}=Q_R+Q_L$ is the conserved total charge of the backscattering model which cancels both from the current and the bias. We thus find that 
\begin{equation}
I(V,T,T_{K},M)=\frac{1}{g_b}\, I_{b}\left(\frac{V}{g_{b}},T,T_{K},g_{b}\right).
\label{eq:currentresult}
\end{equation}
where $I_b$ is calculable using exact results \cite{FendleyPRL95,FLeS1996}.  
The differential conductance is given by the derivative $G= \frac{\partial I}{\partial V}$, providing the combined $T,V$ dependence announced in the introduction.  A further exact result on the BSG model \cite{FendleyNoisePRL95,weiss1999quantum}, combined with the $\Delta Q$ to $\Delta Q_{RL}$ mapping, gives the zero-temperature, zero-frequency shot noise as \cite{Schottkyfootnote} 
\begin{equation}
P=-\frac{e}{1-g_{b}}(VG-I).
\label{eq:noise}\end{equation}
Equations \eqref{eq:currentresult} and \eqref{eq:noise} are our key exact transport results. 

Before moving on to illustrating our exact results, we pause to highlight two key asymptotic features revealed by the Toulouse limit. The first of these concerns the conductance, which in the asymptotic  $T,\ eV\ll T_K$ regime is characterised by a universal scaling function of the ratio $eV/T$. Combining our mapping with sine-Gordon results \cite{weiss1999quantum} we find that $\delta G =G_0-G$ behaves as
\begin{equation}
\frac{\delta G(T,V)}{\delta G(T,0)}=\frac{d}{dx}\,\frac{\sinh(\pi x)}{\pi\Gamma^{2}(g_{b}^{-1})}\left|\Gamma\left(\frac{1}{g_{b}}+ix\right)\right|^{2}
\label{eq:scalingfn}
\end{equation}
where $x=\frac{eV}{2\pi g_{b}T}$ and $\Gamma$ is the gamma function. This scaling function approaches unity for $x\rightarrow 0$ by construction, and for $x\!\gg\! 1$ it is linear with slope $(\frac{2}{g_{b}}-1)/\Gamma^{2}(g_{b}^{-1})$ when plotted against $|x|^{\frac{2}{g_{b}}-2}$. 
The potential use  of Eq.~\eqref{eq:scalingfn} is similar to scaling functions \cite{GrabertWeiss,*FisherDorsey,*vonDelftAP99} pivotal in experimental demonstrations of the two-channel Kondo effect \cite{RalphPRL94,*potok2007observation} or Luttinger liquid behaviour in carbon nanotubes \cite{bockrath1999luttinger}.

The second feature, with a direct noise signature, is the emergence of a striking fractionally quantised charge $e^*$ in the current corrections $\delta I$ near the Kondo fixed point, leading to a fractional Fano factor \cite{BlanterButtiker,*BeenSchon} $F=\frac{P}{2e\delta I}=\frac{e^*}{e}$ due to the events behind $\delta I$ being rare and thus uncorrelated. 
To find $e^*$, we note that for Eq.~\eqref{eq:Luttwire}, it is well known \cite{KaneFisherFraccharge} that at high energies when backscattering is weak, the rare backscattering events are in units of $e^*_h=g_be$;  and for 
low energies when  backscattering becomes strong, its weak correction $\delta I_b$ (i.e., weak tunneling between the two subsystems that emerge due to strong backscattering) is via rare charge $e$ events. 
Since the relation between $\partial_t\Delta Q$ and $\partial_t\Delta Q_{RL}$ is linear, the same high-to-low energy ratio $\frac{e^*_h}{e^*_l}=g_b$ should arise in the topological Kondo effect. At high energies, $\Delta Q$ changes due to electron tunneling, $e^*_h=e$. Therefore, we find that at low energies a fractional charge $e^*=eg_b^{-1}=\frac{2(M-1)}{M}e$ emerges, characterising the weak backscattering contribution $\delta I$ that $\delta I_b$ translates into. The same result is recovered by calculating the Fano factor from Eq.~\eqref{eq:noise}. This charge quantisation, though consistent with asymptotic results on current-current correlation functions \cite{zazunov2014transport}, has not so far received attention. Viewing topological Kondo transport through the Toulouse limit rendered $e^*$ manifest. Note that the value of $e^*$ is characteristic of the topological Kondo effect, and is distinct from the quantised charges noted in other Kondo related systems \cite{Kondonoise}.

\begin{figure}
\includegraphics[trim=0 20 0 0,clip,width=\columnwidth]{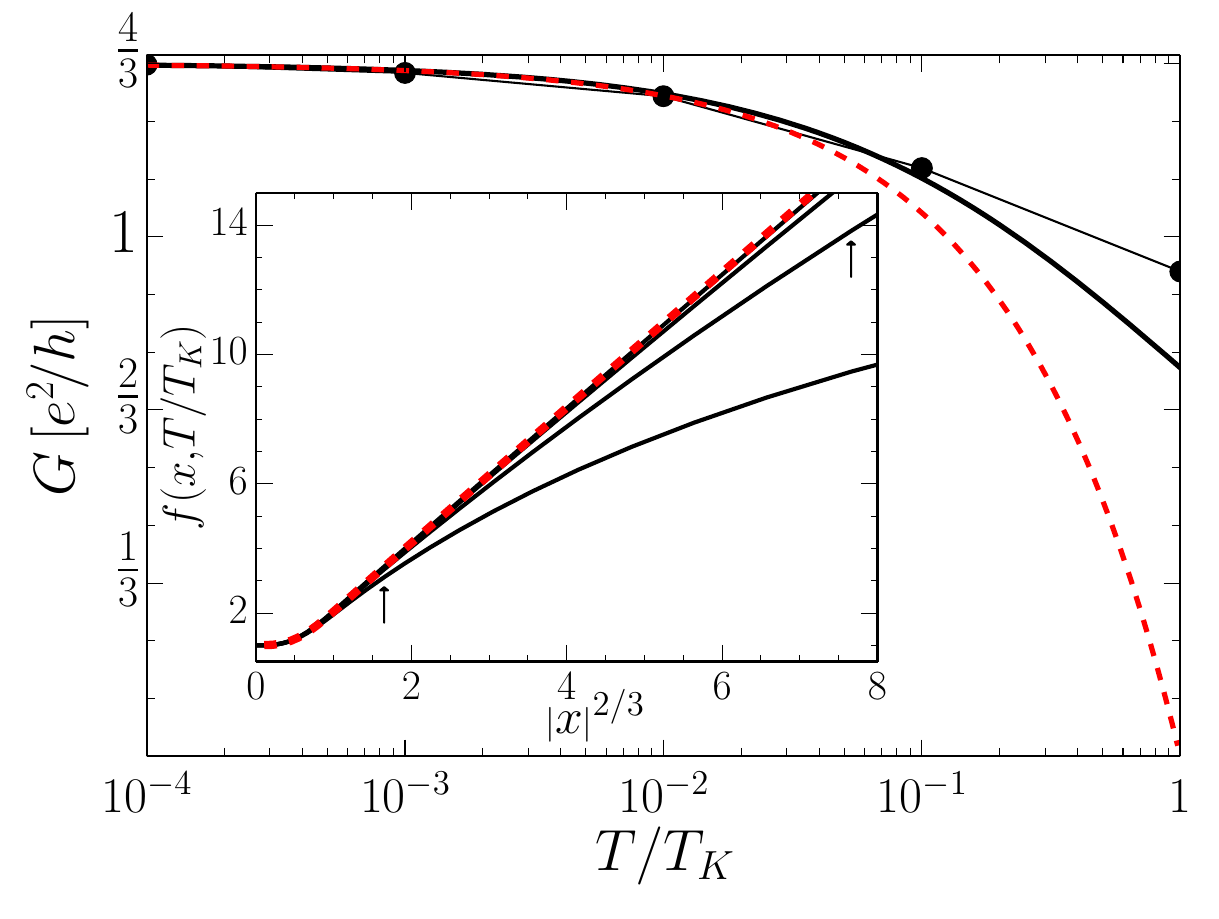}
\caption{Conductance of an $M=3$ device calculated from the Toulouse limit. Main figure: linear conductance (solid line) compared to numerical renormalisation group results \cite{GalpinPRB14} for an isotropic device (dots) and to power law asymptotics (dashed). Inset: the conductance correction $f(x,T/T_K)=\delta G(x,T/T_K)/\delta G(0,T/T_K)$ for temperatures $T=10^{-p}T_K$ with integer $p=2,\ldots,5$ (solid lines) compared to the scaling function in Eq.~\eqref{eq:scalingfn} (dashed). With increasing $p$ the scaling function is followed for an increasing range of $x$. The arrows indicate $eV=0.1T_K$ (for $p=2,3$) beyond which the BSG description breaks down for isotropic systems
and is expected to be replaced by more pronounced departures from scaling.}
\label{fig:Gcurves}
\end{figure}

Finally, we turn to an exact calculation using Eq.~\eqref{eq:currentresult}, focusing on the conductance in the minimal,  $M=3$ lead geometry. In this case, the Luttinger parameter $g_b=\frac{3}{4}$ which allows us to use expressions in Ref.~\onlinecite{FLeS1996} developed for $g_b=1-\frac{1}{n}$ with $n>1$ integer. In terms of $t=\frac{T}{T_K}$ and $v=\frac{eV}{T_K}$ the conductance is  $G(t,v)=\frac{e^{2}}{hg_{b}}\left[1-t\frac{\partial}{\partial v}i(t,\frac{v}{g_{b}})\right]$ where
\begin{multline}
i(t,v)=\frac{3}{2}\int_{-\infty}^{\infty}\frac{d\theta}{\cosh^{2}[\theta+\ln(t)]}\\ \times\ln\left(\frac{1+e^{3v/2t-\varepsilon_{+}(\theta)}}{1+e^{-3v/2t-\varepsilon_{+}(\theta)}}\frac{1+e^{-3v/2t-\varepsilon_{+}(\infty)}}{1+e^{3v/2t-\varepsilon_{+}(\infty)}}\right) .
\end{multline}
Here $\varepsilon_+(\theta)$ is the energy of kinks in the BSG model and $\theta$ is the rapidity. The kink energy $\varepsilon_+(\theta)$ follows from the thermodynamic Bethe ansatz in the form of three coupled integral equations \cite{FLeS1996}, which we solve numerically. 
Our results are shown in Fig.~\ref{fig:Gcurves}. To validate our approach, we first focus on the $V\rightarrow0$ limit where we can compare to numerical renormalisation group results for an isotropic device \cite{GalpinPRB14}. We find that, even for such isotropic systems, the Toulouse limit excellently captures the behaviour in the universal regime with significant deviations appearing only for $T\gtrsim 0.1 T_K$. Next we turn to the combined $T$ and $V$ dependence, and in particular to studying the  emergence of the universal scaling Eq.~\eqref{eq:scalingfn} and the behaviour outside the scaling regime. As the power law begins to break down at $T,eV\gtrsim 0.01 T_K$, scaling is expected to hold until $x^*\sim 2\times10^{p-3}$ for conductance data at $T=10^{-p}T_K$. This is consistent with our exact results (Fig.~\ref{fig:Gcurves} inset) showing the emergence of scaling for $|x|\ll x^*$ and informing on the subsequent deviations.

In conclusion, we have identified a Toulouse limit of the topological Kondo  problem and described how it can be exploited to get exact results and novel physical insights on the universal regime of nonequlibrium transport. In addition to its utility in informing experiments aimed at probing the nonlocality of topological qubits in mesoscopic Majorana devices, our work may open a number of new directions for theoretical research. These may include leveraging the Toulouse limit and available BSG results to obtain a range of novel static and 
dynamical features of the topological Kondo effect, or extending our results to include, e.g., Majorana-Majorana couplings or ``Zeeman terms" in the Kondo language. Using the exponential control over the latter one can study  \mbox{SO($M$) $\rightarrow$ SO($M-2$)} crossovers \cite{ABET1,GalpinPRB14,ABET2} where we anticipate that another effective charge $e^*_{Z}=\frac{2e}{M}$ may emerge and be observable in the Fano factor \cite{BBT2}.

I thank Nigel Cooper for helpful comments on the manuscript and Paul Fendley for a useful conversation.
This research was supported by the Royal Society, the EPSRC grant EP/M02444X/1, and the ERC Starting Grant No. 678795 TopInSy.


\begin{thebibliography}{65}
\bibitem{Toulpertfn} The scaling $\delta G\!\sim\! E^{2/g_b-2}$ is expected to be robust against  deviations from the Toulouse limit: physically, given weak bare $\lambda$, these correspond to moving the system closer to isotropy, the reference point of our comparison.
\end{thebibliography}

%

\end{document}